 \documentclass[aps,prl,amsmath,amssymb,superscriptaddress,showpacs,showkeys,twocolumn]{revtex4-1}
\usepackage{graphicx}
\usepackage{dcolumn}
\usepackage{bm}
\usepackage{siunitx}
\usepackage[utf8]{inputenc}
\usepackage[english]{babel}
\usepackage{natbib}
\usepackage{mathrsfs}
\usepackage{enumerate, mathtools, braket} 
\usepackage{hyperref}

\usepackage{CJK}

\begin{document}
\begin{CJK*}{UTF8}{gbsn} %

\title{Coherent Quantum Interconnection between On-Demand Quantum Dot Single Photons and a Resonant Atomic Quantum Memory} 

\author{Guo-Dong~Cui(崔国栋)}
\altaffiliation{ G.-D.~Cui, L.~Schweickert,  and K.~D.~J\"ons contributed equally to this work.}
\affiliation{Department of Physics \& Astronomy, Stony Brook University, Stony Brook, NY 11794-3800, USA}
\author{Lucas~Schweickert}
\altaffiliation{ G.-D.~Cui, L.~Schweickert,  and K.~D.~J\"ons contributed equally to this work.}
\affiliation{Department of Applied Physics, Royal Institute of Technology, Albanova University Centre, Roslagstullsbacken 21, 106 91 Stockholm, Sweden}%
\author{Klaus~D.~J\"ons}
\email[Corresponding author: ]{klaus.joens@uni-paderborn.de}
\affiliation{Department of Applied Physics, Royal Institute of Technology, Albanova University Centre, Roslagstullsbacken 21, 106 91 Stockholm, Sweden}%
\affiliation{Institute for Photonic Quantum Systems (PhoQS), Center for Optoelectronics and Photonics Paderborn (CeOPP), and Department of Physics, Paderborn University, 33098 Paderborn, Germany}
\altaffiliation{ G.-D.~Cui, L.~Schweickert,  and K.~D.~J\"ons contributed equally to this work.}
\author{Mehdi~Namazi}
\affiliation{Department of Physics \& Astronomy, Stony Brook University, Stony Brook, NY 11794-3800, USA}
\author{Thomas~Lettner}
\author{Katharina~D.~Zeuner}
\author{Lara~Scavuzzo~Monta\~{n}a}
\affiliation{Department of Applied Physics, Royal Institute of Technology, Albanova University Centre, Roslagstullsbacken 21, 106 91 Stockholm, Sweden}%
\author{Saimon~Filipe~Covre~da~Silva} 
\author{Marcus~Reindl}
\author{Huiying~Huang(黄荟颖)}
\affiliation{Institute of Semiconductor and Solid State Physics, Johannes Kepler University Linz, 4040, Austria}
\author{Rinaldo~Trotta}
\affiliation{Institute of Semiconductor and Solid State Physics, Johannes Kepler University Linz, 4040, Austria}
\affiliation{Dipartimento di Fisica, Sapienza Universit\`a di Roma, Piazzale A. Moro 1, I-00185 Roma, Italy}
\author{Armando~Rastelli}
\affiliation{Institute of Semiconductor and Solid State Physics, Johannes Kepler University Linz, 4040, Austria}
\author{Val~Zwiller}
\affiliation{Department of Applied Physics, Royal Institute of Technology, Albanova University Centre, Roslagstullsbacken 21, 106 91 Stockholm, Sweden}%
\author{Eden~Figueroa}
\email[Corresponding author: ]{eden.figueroa@stonybrook.edu}
\affiliation{Department of Physics \& Astronomy, Stony Brook University, Stony Brook, NY 11794-3800, USA}
\affiliation{Brookhaven National Laboratory, Upton, NY, 11973, USA}
\date{\today}

\begin{abstract}
Long-range quantum communication requires the development of in-out light-matter interfaces to achieve a quantum advantage in entanglement distribution. Ideally, these quantum interconnections should be as fast as possible to achieve high-rate entangled qubits distribution. Here, we demonstrate the coherent quanta exchange between single photons generated on-demand from a GaAs quantum dot and atomic ensemble in a $^{87}$Rb vapor quantum memory. Through an open quantum system analysis, we demonstrate the mapping between the quantized electric field of photons and the coherence of the atomic ensemble. Our results play a pivotal role in understanding quantum light-matter interactions at the short time scales required to build fast hybrid quantum networks. 




\end{abstract}

\maketitle
\end{CJK*}

The main goal of quantum communication is to transfer quantum states over arbitrarily long distances in a quantum network. Several proposals exist to use quantum repeater schemes to generate high entanglement rates between remote nodes. An attractive approach to achieve the latter is the use of high-rate entangled photon pair sources \cite{Senellart.Solomon.ea:2017} interfaced with fast quantum memories, capable of receiving and storing entangled states at high speed \cite{Lloyd.Shahriar.ea:2001}. A possible pathway to implement this scheme is the creation of hybrid quantum networks, in which high repetition rate, quantum dot-based sources of on-demand entangled photon pairs~\cite{Muller.Bounouar.ea:2014,Huber.Reindl.ea:2017, Chen.Zopf.ea:2018,Liu.Su.ea:2019,Wang.Hu.ea:2019
} with high purity ~\cite{Schweickert.Joens.ea:2018, Hanschke.Fischer.ea:2018
}, and indistinguishability ~\cite{Somaschi.Giesz.ea:2016, Ding.He.ea:2016, Tomm.Javadi.ea:2021}, are interfaced with fast low-noise, high-fidelity atomic vapor quantum memories~\cite{Namazi.Kupchak.ea:2017,Wolters.Buser.ea:2017,Kaczmarek.Ledingham.ea:2018, Bruser.Treutlein.ea:2022, Finkelstein.Firstenberg.ea:2018, Michelberger.Walmsley.ea:2015, Reim.Walmsley.ea:2010}.

\begin{figure*}[ht]
\includegraphics[width=\textwidth]{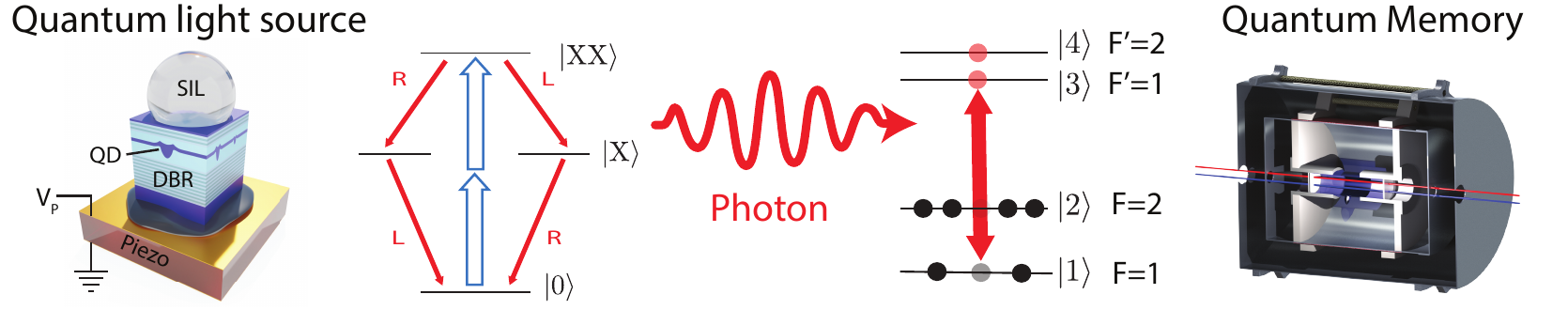}
\caption{Hybrid quantum light-matter interfaces. 
(Left) A tunable solid-state quantum dot photon source. The quantum dot emits entangled photon pairs with different frequencies via its biexciton-exciton cascade. The biexciton transition is strain-tuned to emit at atomic transition wavelengths. 
(Right) A $^{87}\text{Rb}$-based room temperature quantum light-matter interface with $D_1$ line ground states $F=1,2$ (labeled as $\ket{1}$, $\ket{2}$) coupled to two excited states $F'=1,2$ (labeled as $\ket{3}$, $\ket{4}$). } \label{fig:schematic}
\end{figure*}
%

The key challenge to constructing these fast hybrid quantum networks~\cite{Schimpf.Reindl.ea:2021, Neuwirth.Basset.ea:2021} remains to demonstrate the quanta exchange between large-bandwidth single photons generated in quantum dots and atomic quantum memories. This mapping between quantized electric fields and atomic coherence has only been shown for narrow bandwidth photons ~\cite{Matsukevich.Kuzmich:2004,Eisaman.Andre.ea:2005}. A series of experiments have demonstrated the interaction of quantum dot-generated single photons with atomic vapor ~\cite{Akopian.Wang.ea:2011, Siyushev.Stein.ea:2014, Jahn.Munsch.ea:2015, Trotta.Martin-Sanchez.ea:2016, Portalupi.Widmann.ea:2016, Vural.Portalupi.ea:2018, Kroh.Benson.ea:2019}. However, all these experiments rely on the dispersive off-resonant slow-down effect generated by placing a photon envelope in between two absorption resonances \cite{Camacho.Howell.ea:2006}. This effect can be explained using a steady-state semi-classical framework, without the need for a density matrix treatment \cite{Fleischhauer.Lukin:2000}. 

In this paper, we experimentally demonstrate and theoretically verify the coherent quanta exchange between the single-photon field generated by a quantum dot and the atomic ensemble of a room-temperature quantum memory. As we could not directly probe the state of the ensemble during the light-matter interaction, we developed an indirect method to obtain the quantized atomic response. Fig.~\ref{fig:schematic} shows a schematic of the experimental concept, where a solid-state quantum light source emits pairs of photons and one is interfaced with an atomic quantum memory.

We first produce on-demand photon pairs from a tunable GaAs quantum dot, where one photon of the pair is tuned to the $^{87}\text{Rb}$ $D_1$ line $F=1\leftrightarrow F'=1$ transition. Secondly, we interact the photons with an atomic ensemble following a resonant excitation process. Lastly, we compare the spectral manipulation of the photon's temporal wave function after atomic interaction, to a Maxwell-Bloch equations simulation, from which we obtain the spatio-temporal evolution of the density matrix elements describing the coherent mapping between the quantized electrical field and the atomic populations and coherence. The complete hybrid quantum network used for these experiments is depicted in Fig. SM1.



The photon generation follows the following procedure. The picosecond(ps) pulses of an optical parametric oscillator laser are shaped into 10 ps long pulses to be used in a two-photon resonant excitation scheme~\cite{Brunner.Abstreiter.ea:1994,Stufler.Machnikowski.ea:2006}. This procedure excites the quantum dot which is located in a closed-cycle helium cryostat (5 K sample temperature). The excitation laser is filtered using cross-polarization as well as a narrow-band notch filter. The cross-polarization is aligned to the principle axis of the quantum dot to reject one of the exciton fine-structure components. The exciton and biexciton transition are separated via a transmission spectrometer with an end-to-end efficiency of 70\,\%. 

\begin{figure*}[htb] 
\includegraphics[width=1\textwidth]{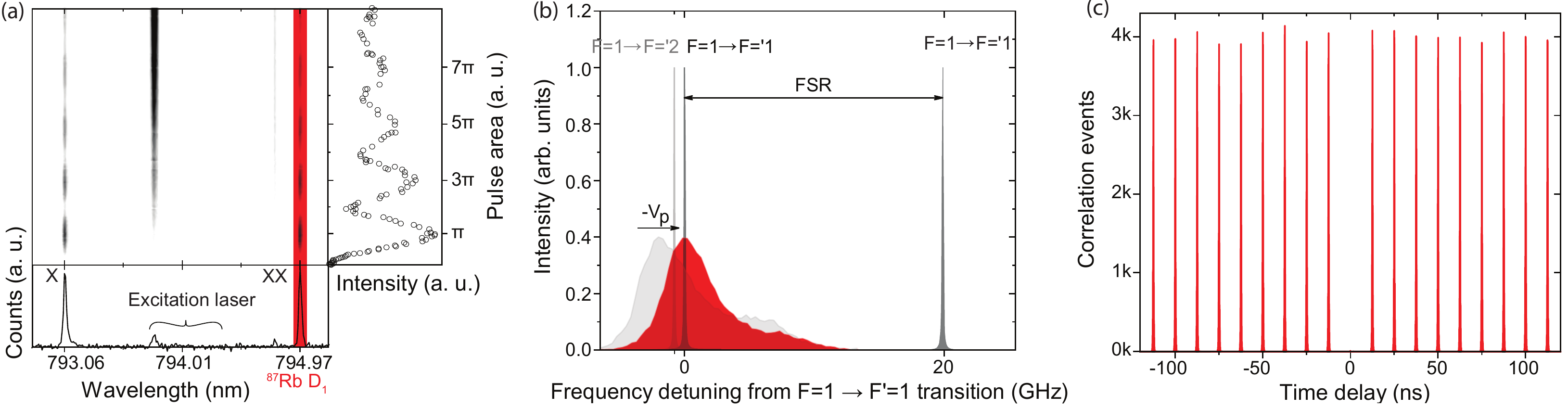}
\caption{Quantum dot light source characterization. (a) Color-coded photoluminescence spectra of the biexciton-exciton cascade under resonant two-photon excitation as a function of the excitation pulse area, revealing Rabi-oscillations. The red vertical line indicates the rubidium D$_1$ transitions.  (b) High-resolution spectroscopy with a tunable Fabry-P\'{e}rot interferometer to precisely measure the XX transition of the quantum dot with respect to the rubidium $F = 1 \leftrightarrow F' = 1$ transition.  By fine-adjusting the piezo voltage (V$_P$) the XX line is brought from off-resonance (gray area) into resonance (red area).  (c) Second-order intensity auto-correlation measurement of the XX transition under coherent $\pi$-pulse excitation, shows g$^{(2)}(0)=0.0018\pm0.0001$.\label{fig:qd}}
\end{figure*}

To achieve the required tunability, we use GaAs quantum dots grown by molecular beam epitaxy with an s-shell biexciton (XX) resonance designed to be close to the D$_1$ line of rubidium at 795\,nm. We can precisely tune the XX emission wavelength by applying stress to the quantum dot structure via a piezo-electric substrate. The quantum dot is excited with a two-photon resonant process in a confocal microscopy setup. Fig.~\ref{fig:qd} (a) shows color-coded photoluminescence spectra of the biexciton-exciton cascade as a function of the excitation laser pulse area. The Rabi oscillations verify the coherent control of the quantum dot system and all further measurements are performed with an excitation laser pulse area corresponding to a $\pi$-pulse. Fig. SM2 shows the reversibility of the tuning process.

 To verify the correct tuning with respect to the desired rubidium absorption lines we employ a high-resolution photoluminescence spectroscopy (HRPL) setup consisting of a tunable fiber-coupled Fabry-P\'{e}rot interferometer with a resolution of 70 MHz and free spectral range (FSR) of 20 GHz. Fig.~\ref{fig:qd} (b) displays our HRPL measurements. Two FSR orders of a laser locked to the $F = 1 \leftrightarrow F' = 1$ and one FSR order of a laser locked to the $F = 1 \leftrightarrow F' = 2$ 
are shown as a reference to determine the absolute detuning of the quantum dot XX transition to the desired Rb resonance. Two biexciton HRPL spectra are shown for different piezo tuning ($\Delta V_p = 11$ V), highlighting our achieved tuning precision. For clarity, only one spectral order of the XX emission is plotted. The tail at the right (low energy side) of the XX transition corresponds to the phonon sideband. 

The biexcitonic single photon emission of our coherently-driven quantum dot is sent to a Hanbury Brown and Twiss setup to measure the spectral correlations between subsequently emitted photons using time tagging. Fig.~\ref{fig:qd} (c) depicts the histogram after correlating the time-tagged data to reassemble a pulsed second-order intensity auto-correlation $g^{(2)}(\Delta \tau)$-function. The distance between the side peaks corresponds to the laser repetition rate of ~80 MHz. As a result a g$^{(2)}$($\tau$) correlation function is measured with a final g$^{(2)}(0)=0.0018\pm0.0001$, showing almost perfect single photon emission. Several quantum dots have been similarly characterized, showing comparable optical properties, including: (i) less than a factor of 2 away from the Fourier limit, (ii) XX lifetimes of approximately 130 ps, (iii) wavelength tunability of 0.5 nm, (iv) extremely low multi-photon emission, and (v) small fine-structure splitting.

\begin{figure*}[htb]
	\includegraphics[width = .9\textwidth]{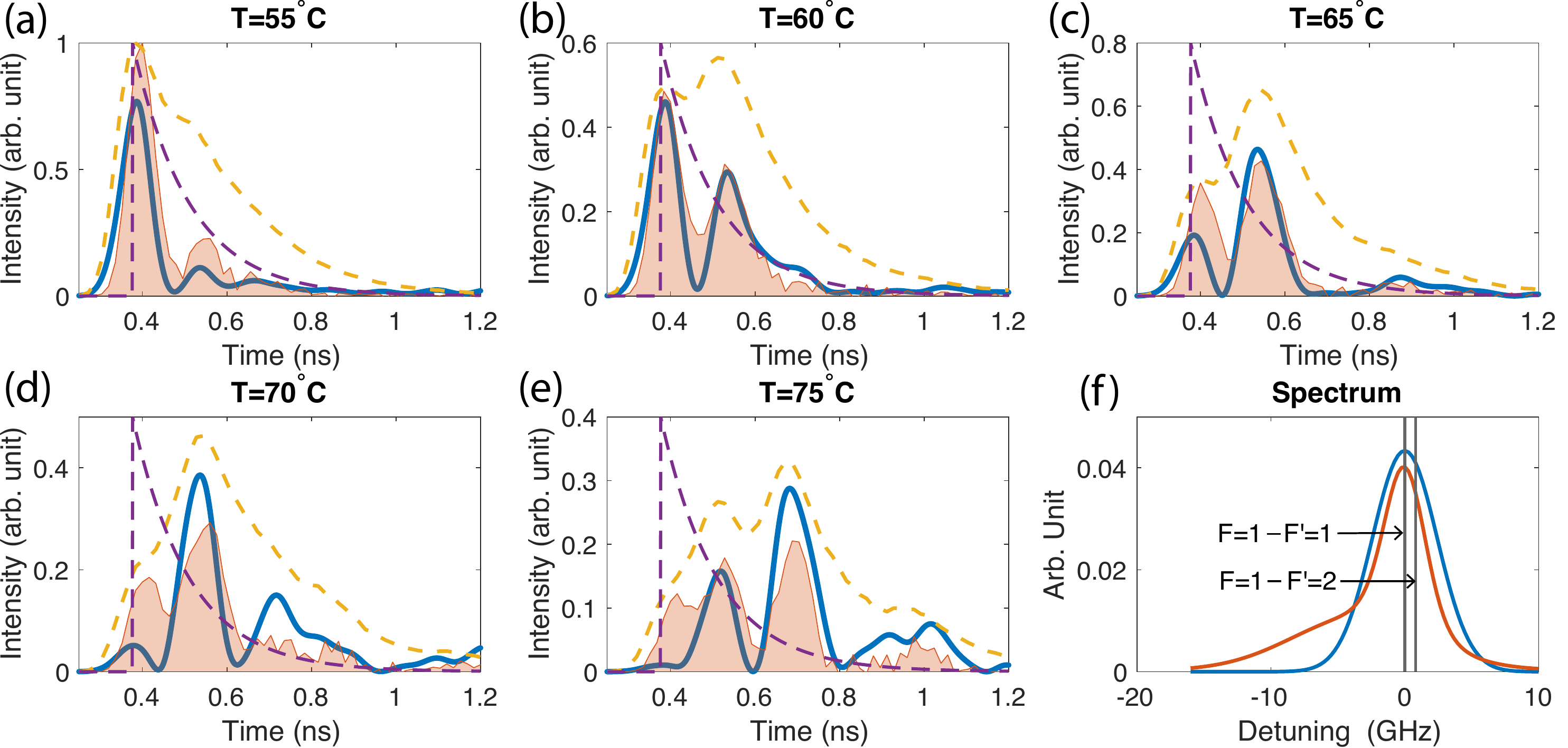}
	\caption{(a-e): Benchmarking the simulation (blue line) against deconvoluted experimental data (red area) for different cell temperatures. All simulations are conducted with the same parameters except for different ensemble temperatures to achieve a global fit with the data. We used an effective Gaussian temporal profile with 77.56 ps FWHM as input. Deconvoluted data at each temperature is obtained from the measured histogram (yellow dashed line) deconvoluted with the exponential decay (purple dashed line) and is normalized to the 55$^\circ$C data. 
	(f) The effective Gaussian spectrum used in the simulation (blue line) and the measured spectrum (red line). }
	\label{fig:tempseries}
\end{figure*}

We collect the quantum dot photons with polarization maintaining fiber serving as a quantum communication channel towards a room temperature quantum memory \cite{Namazi.Kupchak.ea:2017}. The quantum memory is based upon a warm $^{87}$Rb vapor cell at approximately $60^{\circ}\,$C, containing Kr buffer gas and serves as the atomic medium to transform the short single photons. The memory-transmitted photons are collected in a superconducting nanowire single-photon detector (SNSPD) in order to study the resonant dynamics of the quantum light-matter interface. The dashed yellow lines in Fig. \ref{fig:tempseries} are the histograms collected for the transmitted photons through the memory for different temperatures. In order to account for the uncertainty in the emission time of the photons, described by the Wigner-Weisskopf theory \cite{Weisskopf.Wigner.1930} applied to the bi-exciton transition, we de-convolute these experimental histograms with a 134 ps lifetime exponential decay form. This lifetime is obtained from the measured photon arrival time histogram without atomic interaction [Fig. \ref{fig:tempseries}, dashed purple lines]. The results of these procedures are shown as red histograms in Fig. \ref{fig:tempseries} and can be interpreted as the inherent temporal evolution of the short photons created in the quantum dots as they transverse a resonant atomic medium. 

As the purpose of this work is to go beyond the semi-classical description of the light-matter interface, we now construct a quantized model of the interaction between temporally short single photons and coherently prepared multi-level atomic transitions. This model will allow us then to explore the quantum state of the atoms during and after the photon propagation. The input temporal envelope of the photons impinging on the quantum memory used in our model is derived from the measurements of the photon spectrum width, which is presented in Fig. SM3. This allows us to use an effective Gaussian temporal envelope ${\mathcal E}_\text{eff}(t)$ with full width at half maximum (FWHM) 77.56 ps corresponding to the inverse Fourier transform of the measured spectral FWHM 5.69 GHz that accounts for the natural linewidth, the spectral diffusion, and the phonon sideband (details in SM Section IV, Fig. SM3).

We model the interaction of this single photon temporal envelope with four-level atoms (defined in Fig. \ref{fig:schematic}) using a rotating wave approximation Hamiltonian:
\begin{equation}\label{eq_hamiltonian}
\begin{aligned}
\hat{H}(&z,t) = \hbar \omega_{21} \hat \sigma_{22} -\hbar\Delta_p \hat \sigma_{33} +\hbar(\omega_{43}-\Delta_p) \hat \sigma_{44} \\ &
-i \left(d_{31} \hat \sigma_{31} + d_{41} \hat \sigma_{41} + d_{32} \hat \sigma_{32} + d_{42} \hat \sigma_{42}s  \right) \hat {\mathcal E}(z,t)  + H.c. 
\end{aligned}
\end{equation}
Here, $\hat \sigma_{ij} (i,j = 1,2,3,4)$ are the atomic operators, $d_{ij}$ are the dipole moments, $\Delta_p=\omega_p-\omega_{31}$
is the detuning between the photon carrier frequency and the $F=1\leftrightarrow F'=1$ transition, 
and $\hat {\mathcal E}(z,t)$ is the slow envelope of the single photon field (positive frequency part).
We obtain the spatio-temporal evolution of quantum dynamics of both the atomic ensemble and the photon field by solving the coupled Maxwell-Bloch equations.  Our simulations account for the full physical system, which in the D$_1$ line in $^{87}\textrm{Rb}$ atoms contains two hyperfine ground states ($F=1,2$) and  two excited states ($F'=1,2$). We use a Doppler width of 500\,MHz and a pressure broadening of 300\,MHz to account for the two broadened manifold resonances (separated by 6.834\,GHz), each with a FWHM resonance of 800\,MHz. Our model is described in detail in SM Section V.




\begin{figure}[hb]
\includegraphics[width=\columnwidth]{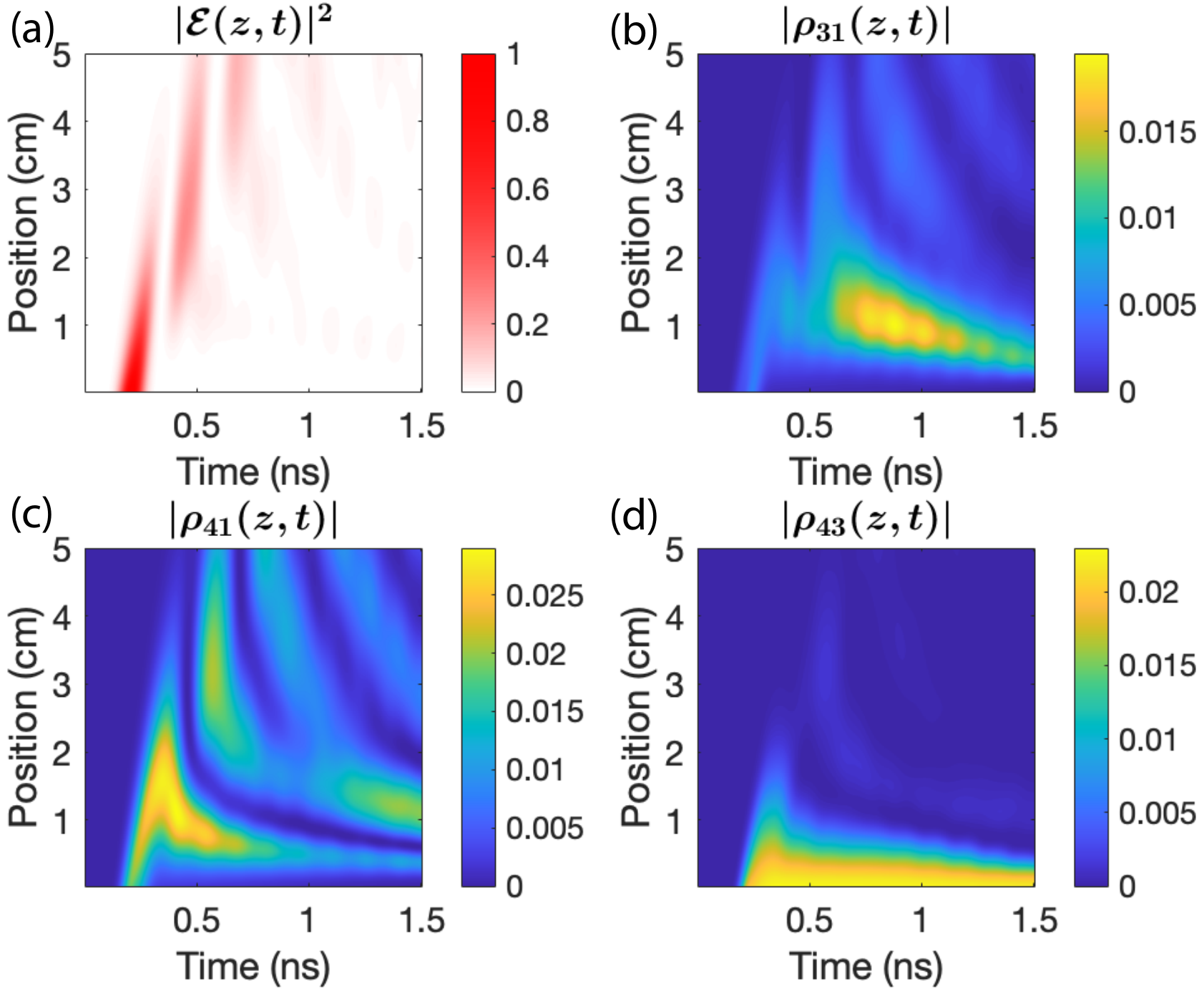}
\caption{Simulated spatio-temporal distribution of (a) photon field intensity, (b,c) atomic density matrix off-diagonal elements corresponding to the coherences between excited state $\ket{3}$, $\ket{4}$ and the ground state $\ket{1}$, (d) the coherence between the two excited states. The simulation shows the evolution in a time span of 1.5 ns, across the 5 cm rubidium cell at 75 $^\circ{C}$.}
\label{fig:coherence_photon}
\end{figure}

The comparison between this quantized model and the measured histograms is shown in Fig.~\ref{fig:tempseries}. 
The solid blue lines in Fig. \ref{fig:tempseries} (a-e) demonstrate the simulated intensity $|{\mathcal E}(z=5\text{ cm},t)|^2$, for the initial temporal envelope ${\mathcal E}_\text{eff}(t)$, after propagation through the 5 cm cell. 
We emphasize that all parameters in this model are the same except for the various ensemble temperatures.
The global agreement between the numerical model and the experimental data enables us to make inferences about additional physical phenomena occurring in fast resonant quantum light-matter interactions.
More details on our benchmarking between simulation and experiment are explained in SM Section VI.

By simulating the evolution of the relevant density matrix elements in time and space as the resonant pulse propagates through the cell [Fig. \ref{fig:coherence_photon}], we observe novel dynamics as compared to the off-resonant interactions reported previously~\cite{Akopian.Wang.ea:2011, Trotta.Martin-Sanchez.ea:2016,Vural.Portalupi.ea:2018}. The simulated evolution of the resonant photon field $|\mathcal{E}(z,t)|^2$ shows the creation of several time-dependent oscillations \cite{Kallmann.Hartmann.ea.1999}. These are also present in the density matrix element $\rho_{31}(z,t)$, corresponding to the polarizability of the resonant transition $F = 1 \leftrightarrow F' = 1$. At deeper propagation depths, oppositely phased oscillations develop between $|\mathcal{E}(z,t)|^2$ and $\rho_{31}(z,t)$, indicating the formation of a light-matter polariton. Analogously, a polariton forms between $|\mathcal{E}(z,t)|^2$ and $\rho_{41}(z,t)$, corresponding to $F = 1 \leftrightarrow F' = 2$ transition.
Additionally, we observe in-phase oscillations between $\rho_{31}(z,t)$ and $\rho_{41}(z,t)$ testifying to the formation of a non-linear coupled system, in the form of an atomic V-scheme. 
Lastly, we observe the formation of an excited coherence $\rho_{43}(z,t)$, which outlasts the original duration of the input pulses by a factor of $\sim{20}$.

\begin{figure}[htb]
	\includegraphics[width = 1\columnwidth]{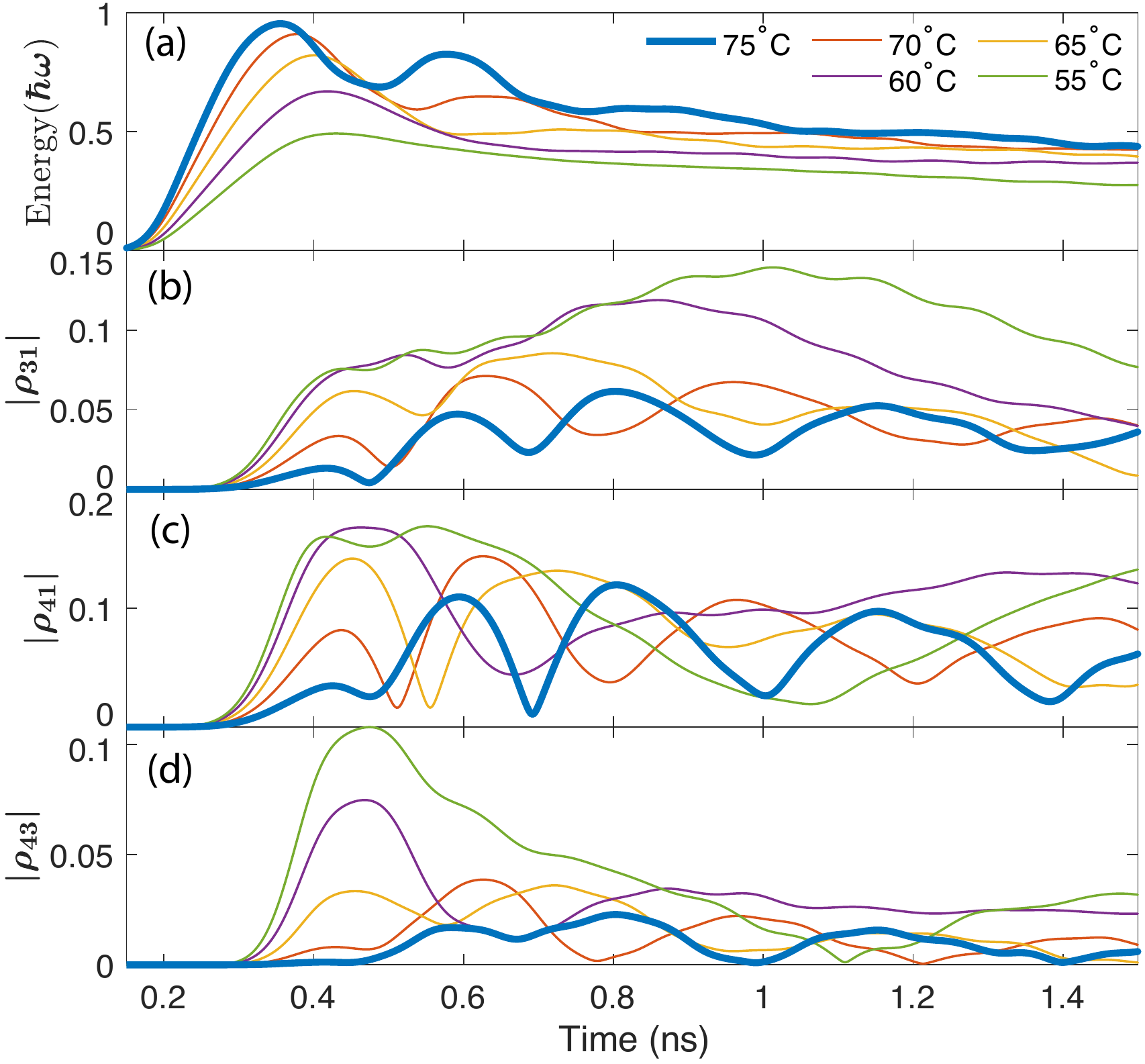}
	\caption{(a) Simulation of ensemble energy accumulation during and after quantum dot photon passes through the vapor cell at various temperatures.
	The accumulated energy is calculated in units of photon quanta.
	The decay rate of the accumulated energy is much slower compared with the free photon dynamics, revealing storage inside the ensemble.
    More simulations can be found in SM.
	(b,c,d) Simulated coherence absolute values at propagation depth $z=5$ cm.}
	\label{fig:ensemble_energy}
\end{figure}

To highlight these novel features, we plot the atomic ensemble energy [defined in SM Section VII], and coherence $\rho_{31}(t)$, $\rho_{41}(t)$, $\rho_{43}(t)$ for the largest propagation depth, $z=5$ cm, at different cell temperatures[Fig. \ref{fig:ensemble_energy}]. Here we observe the ensemble energy remains non-zero for times far beyond the original temporal length of the incoming pulse. Our interpretation of these findings is that the ensemble retains the original optical excitation in the form of a coherent superposition of the atomic excited states, akin to what a storage-of-light procedure achieves in the ground states. This is independently verified by the evolution of the exited state coherence $\rho_{43}(t)$. 
Most noticeably, as temperature increases, the coherences $\rho_{31}(t)$ and $\rho_{41}(t)$ become coupled, following the quasi-simulton-like dynamics discovered in \cite{Ogden.Potvliege:2019}. In our case, we use single photons in a superposition of many frequency modes, two of which interact simultaneously with the two coupled atomic transitions, thus creating a superposition of two single-photon polaritons, what we call a quantum quasi-simulton. 

In summary, we have found a novel regime of coherent dynamics in which short single photons from GaAs quantum dot are modified by their propagation in a resonant $^{87}$Rb ensemble. 
We modeled the coherent quanta exchange between the single-photon field and the atomic ensemble, and benchmarked the model experimentally.
The predicted features of the system provide a promising pathway toward developing non-linear photonic quantum devices.
The formation of long-lived atomic coherence in the system could be used to store short photons with a user-defined retrieval time by replacing one of the arms of the V-scheme with a time-modulated classical field. 
Additionally, the quantum quasi-simulton dynamics could be used to construct non-linear photon-photon gates by sending two independent narrower-in-frequency photons in the same V-scheme. 
Naturally, these techniques are of great interest in the context of creating large hybrid quantum networks.
We envision pursuing these topics in future experiments.  



\begin{acknowledgments}
This work was financially supported by the European Research Council (ERC) under the European Union's Horizon 2020 research and innovation programme (SPQRel, Grant Agreement 679183) and the European Union Seventh Framework Program 209 (FP7/2007-2013) under Grant Agreement No. 601126 210 (HANAS) and No. 899814 (Qurope). The JKU group acknowledges the Austrian Science Fund (FWF): P29603 and V. Volobuev, Y. Huo, P. Atkinson, G. Weihs, and B. Pressl for fruitful discussions. 
The Stony Brook group acknowledges funding from the National Science Foundation, grants number PHY-1404398 and PHY-1707919, and the Simons Foundation, grant number SBF241180.
\end{acknowledgments}



\bibliography{reference} 
\bibliographystyle{apsrev4-1} 

\end{document}


\begin{CJK*}{UTF8}{gbsn} %

\title{SUPPLEMENTARY MATERIALS 
\\Coherent Quantum Interconnection between On-Demand Quantum Dot Single Photons and a Resonant Atomic Quantum Memory}


\author{Guo-Dong~Cui(崔国栋)}
\altaffiliation{ G.-D.~Cui, L.~Schweickert,  and K.~D.~J\"ons contributed equally to this work.}
\affiliation{Department of Physics \& Astronomy, Stony Brook University, Stony Brook, NY 11794-3800, USA}
\author{Lucas~Schweickert}
\altaffiliation{ G.-D.~Cui, L.~Schweickert,  and K.~D.~J\"ons contributed equally to this work.}
\affiliation{Department of Applied Physics, Royal Institute of Technology, Albanova University Centre, Roslagstullsbacken 21, 106 91 Stockholm, Sweden}%
\author{Klaus~D.~J\"ons}
\email[Corresponding author: ]{klaus.joens@uni-paderborn.de}
\affiliation{Department of Applied Physics, Royal Institute of Technology, Albanova University Centre, Roslagstullsbacken 21, 106 91 Stockholm, Sweden}%
\affiliation{Institute for Photonic Quantum Systems (PhoQS), Center for Optoelectronics and Photonics Paderborn (CeOPP), and Department of Physics, Paderborn University, 33098 Paderborn, Germany}
\altaffiliation{ G.-D.~Cui, L.~Schweickert,  and K.~D.~J\"ons contributed equally to this work.}
\author{Mehdi~Namazi}
\affiliation{Department of Physics \& Astronomy, Stony Brook University, Stony Brook, NY 11794-3800, USA}
\author{Thomas~Lettner}
\author{Katharina~D.~Zeuner}
\author{Lara~Scavuzzo~Monta\~{n}a}
\affiliation{Department of Applied Physics, Royal Institute of Technology, Albanova University Centre, Roslagstullsbacken 21, 106 91 Stockholm, Sweden}%
\author{Saimon~Filipe~Covre~da~Silva} 
\author{Marcus~Reindl}
\author{Huiying~Huang(黄荟颖)}
\affiliation{Institute of Semiconductor and Solid State Physics, Johannes Kepler University Linz, 4040, Austria}
\author{Rinaldo~Trotta}
\affiliation{Institute of Semiconductor and Solid State Physics, Johannes Kepler University Linz, 4040, Austria}
\affiliation{Dipartimento di Fisica, Sapienza Universit\`a di Roma, Piazzale A. Moro 1, I-00185 Roma, Italy}
\author{Armando~Rastelli}
\affiliation{Institute of Semiconductor and Solid State Physics, Johannes Kepler University Linz, 4040, Austria}
\author{Val~Zwiller}
\affiliation{Department of Applied Physics, Royal Institute of Technology, Albanova University Centre, Roslagstullsbacken 21, 106 91 Stockholm, Sweden}%
\author{Eden~Figueroa}
\email[Corresponding author: ]{eden.figueroa@stonybrook.edu}
\affiliation{Department of Physics \& Astronomy, Stony Brook University, Stony Brook, NY 11794-3800, USA}
\affiliation{Brookhaven National Laboratory, Upton, NY, 11973, USA}

\maketitle
\end{CJK*}

\section{Experimental setup} 
\label{method:setup}

\begin{figure*}[h]
\includegraphics[width=0.65\textwidth]
{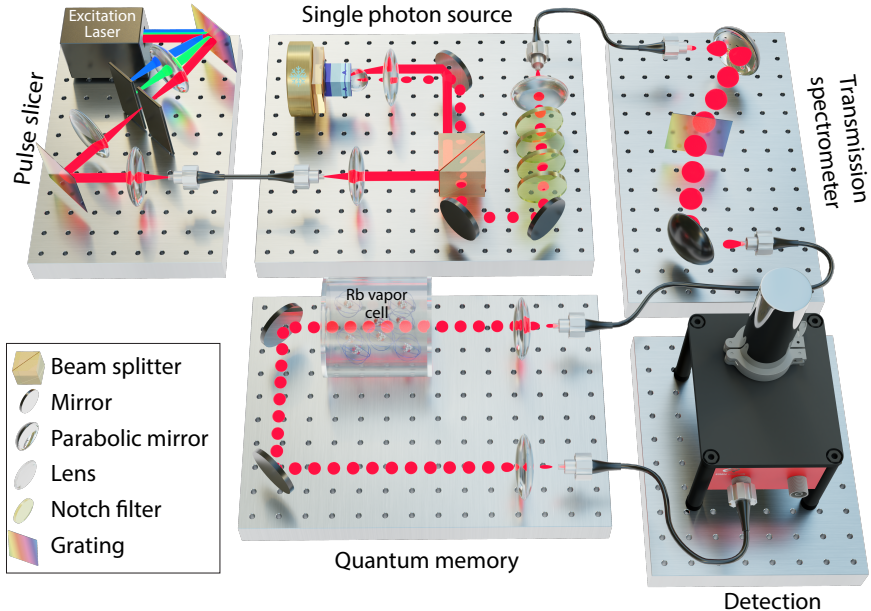}
\caption{Combined setup to interface quantum dot single photons and a rubidium light-matter interface. 
\textbf{Upper left}: Pulsed laser system for single quantum dot excitation.
\textbf{Upper center}
quantum dot single photon production.
\textbf{Upper right}: spectral characterization of the quantum dot photons.
\textbf{Lower center}: quantum memory setup including rubidium room temperature ensemble.
\textbf{Bottom right}: single photon detection after interaction. 
\label{fig:setup}}
\end{figure*}

The experiment setup of our hybrid quantum platform is demonstrated in Fig. \ref{fig:setup}.
It consists of three main parts. 
The first part is the quantum-light source, which generates on-demand pairs of entangled photons, with one photon of the pair specifically tuned to generate quantum fields at the rubidium D$_1$ line. 
In the second part, we shine the temporally short single photons onto a room temperature rubidium ensemble, thereby coherently controlling the temporal wave function of the short single photons by the on-resonance interaction with the rubidium atoms. 
Lastly, after atomic interaction, we detect the photons using a superconducting nanowire single-photon detector (SNSPD).

\section{Quantum dot sample} 
\label{method:sample}
The quantum dot sample was grown by molecular beam epitaxy at the JKU in Linz. First a bottom distributed Bragg reflector made of 9 pairs of $\lambda$/4-thick Al$_{0.95}$Ga$_{0.05}$As (\SI{68.9}{\nano\meter}) and Al$_{0.2}$Ga$_{0.8}$As layers is deposited as a mirror for a $\lambda$-cavity. The quantum dot layer is located at the center of the $\lambda$-cavity made of a $\lambda$/2-thick (\SI{123}{\nano\meter}) layer of Al$_{0.4}$Ga$_{0.6}$As sandwiched between two $\lambda$/4-thick (\SI{59.8}{\nano\meter}) Al$_{0.2}$Ga$_{0.8}$As layers. The quantum dot layer is fabricated by Al-droplet etching~\cite{Heyn.Stemmann.ea:2009,Huo.Witek.ea:2014} on the Al$_{0.4}$Ga$_{0.6}$As layer followed by deposition of \SI{2}{\nano\meter} GaAs. The top mirror for the cavity is made of two pairs of the same material combination as the bottom distributed Bragg reflector. To protect the structure a 4 nm-thick GaAs protective layer covers the final sample. The QD design emission wavelength is centered around $\sim$\,\SI{790}{\nano\meter} and a gradient in the mode position (Q factor of about 50) is generated by stopping the substrate rotation during the top Al$_{0.2}$Ga$_{0.8}$As cavity-layer deposition. The cavity design together with solid immersion lens enhances the collection efficiency by $\sim$\,30 times compared to an unstructured sample.

\section{Piezo-actuator integration} 
\label{method:piezo}

\begin{figure}[ht]
\includegraphics[width=.4\textwidth]
{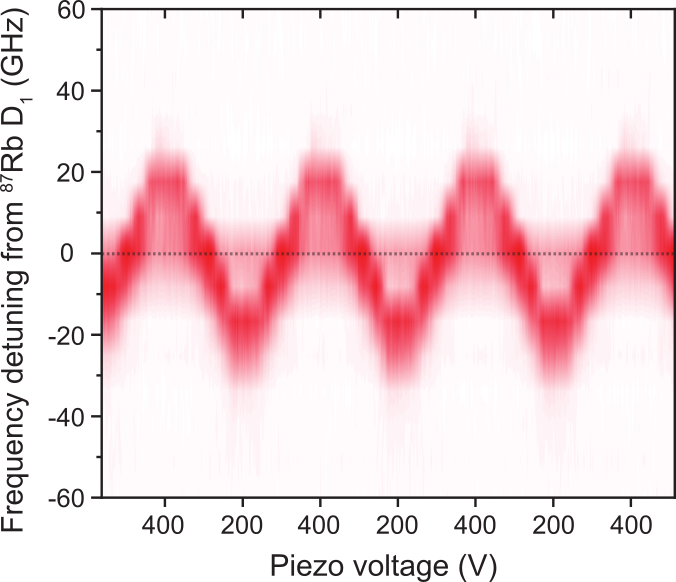}
\caption{Color-coded photoluminescence spectra of the XX transition tuned in and out of resonance of rubidium D$_1$ transition. The horizontal line is a narrow-band cw laser locked to the Rubidium D$_{1}$ line $5S_{1/2} F = 1$ $\rightarrow$ $5P_{1/2} F' = 1$ transition as a reference.
}
\label{fig:SI_QDtuning}
\end{figure} 

To achieve stable and precise energy tuning of our quantum dot emission into resonance with the rubidium $D_1$ $F=1\rightarrow F'=1$ transition we integrate the GaAs quantum dot structure on a gold coated PMN-PT piezoelectric actuator (TRS technologies, thickness of $200\,\si{\micro\meter}$, $\langle 001 \rangle$ orientation). The quantum dot sample is mechanically thinned using diamond-based abrasive films and transferred on the piezo-electric substrate  with a bendable soft tip in pick and place method, taking advantage of electrostatic forces. We use cryogenic epoxy (Stycast, two component resin) resulting in a rigid connection between the sample and the piezo-actuator with good thermal contact. More details on the quantum dot sample transfer on the piezo-electric substrate can be found in Ref.~\onlinecite{Zeuner.Paul.ea:2018}. To verify the reversible emission energy tuning we performed several piezo-actuator voltage sweeps. Fig. 2 in the main text shows the tuning of the quantum dot emission by the application of a DC voltage to the piezoelectric actuator, the emission is compared to a laser locked to the rubidium D$_{1}$ line $5S_{1/2}\,F = 1$ $\rightarrow$ $5P_{1/2}\,F' = 1$ transition. Our device can be repeatably tuned in and out of resonance with a much higher precision than the resolution of the spectrometer [see Fig. \ref{fig:SI_QDtuning}].

\section{Effective Photon Frequency and Temporal Profiles}\label{section:envelope}

\begin{figure}[htb]
\includegraphics[width=.5\textwidth]
{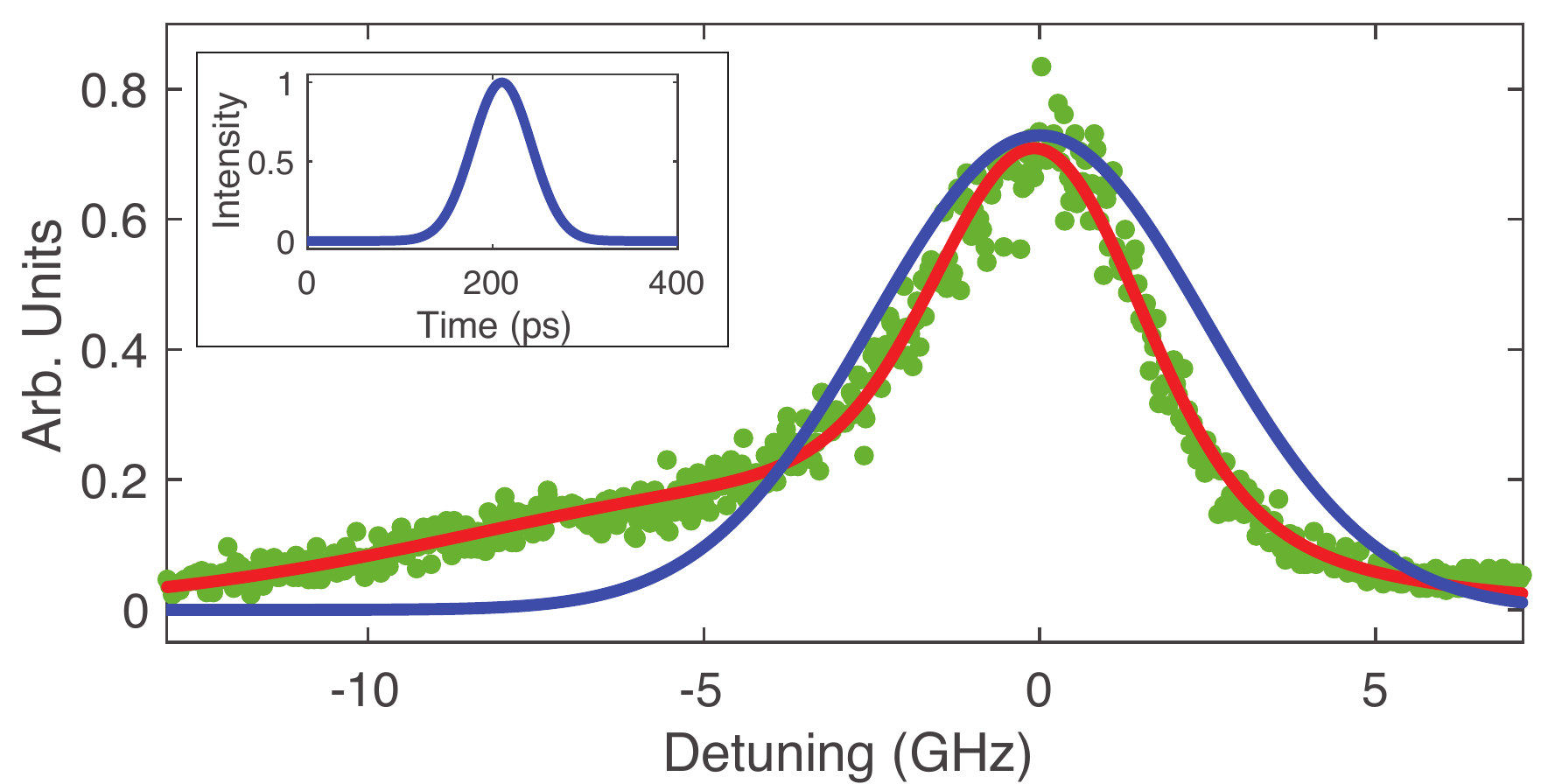}
\caption{Spectrum of quantum dot photon.
The horizontal axis is the detuning of the Fabry-Perot interferometer from $^{87}$Rb $D_1$ transition $F=1 \leftrightarrow F'=1$ line in GHz.
\textbf{Green dots}: transmitted photon histograms through Fabry Perot interferometer at different resonance frequencies.
\textbf{Red curve}: fitted spectrum using two added Voigt intensity profiles.
One of them is the convolution of a Lorentzian with FWHM of 1.18 GHz (corresponding to the 134 ps measured lifetime)  and a Gaussian with a variance of 1.31 GHz (corresponding to spectral diffusion broadening).
The other, centered at -4 GHz away, is the convolution of a Lorentzian with FWHM of 1.19 GHz and a Gaussian with a variance of 4.62 GHz (corresponding to the photon sideband).
\textbf{Blue curve}: the {effective} Gaussian spectrum used to capture the dominant dynamics in our simulation.
The FWHM is fitted to be 5.69 GHz using one Gaussian profile.
\textbf{Insert}: the effective Gaussian intensity profile versus time, that is obtained via the inverse Fourier transformation from the effective Gaussian spectrum.
The FWHM is calculated to be 77.56 ps.
It is used as the input single-photon temporal profile in the quantum photon-atom interaction simulation.
}
\label{fig:lineshape}
\end{figure}

To characterize the spectrum of quantum dot photons with respect to the desired Rb resonance, a high-resolution photoluminescence spectroscopy (HRPL) setup is employed with the help of a tunable fiber-coupled Fabry-P\'{e}rot interferometer that could resolve 70\,MHz and has a free spectral range (FSR) of 20\,GHz.
The measured spectral lineshape of the bi-exciton (XX) photon is presented in Fig. \ref{fig:lineshape}.

In order to perform our simulation of the single-photon dynamics, we use an effective frequency spectrum of the quantum dot single photon as seen by the atomic ensemble. This can be represented by a Gaussian profile, as indicated in the blue profile presented in Fig. \ref{fig:lineshape}.
The effective line-shape width is determined by a Gaussian fit to the measured spectrum. This follows the form:
\begin{equation}
\begin{aligned}
S_\text{eff} (\nu) \propto  e^{- 4\ln(2)\frac{\nu^2}{\Delta\nu^2}}
\end{aligned}
\end{equation}
The full width at half maximum (FWHM) is $\Delta\nu = 5.69$ GHz.
From the inverse Fourier transformation of the effective frequency spectrum, we get the effective temporal profile of the photon in the time domain:
\begin{equation}
\begin{aligned}
\mathcal{I}_\text{eff} (t) \propto |\mathcal{E}_\text{eff} (t)|^2 \propto \left| FT^{-1}\left[\sqrt{S_\text{eff} (\nu)}\right](t)\right|^2 
\propto e^{- 4\ln(2)\frac{t^2}{\Delta{t}^2}}
\label{eq:effective_temporal}
\end{aligned}
\end{equation}
where the photon's FWHM in time is calculated to be $\Delta{t} = 77.56$ ps. This is displayed in the inset of Fig. \ref{fig:lineshape} and it is the input of our master equation simulation to calculate the atomic dynamics. 

\section{Maxwell-Bloch Equations Simulation of Light-Matter Interaction}
We assume that the field transverses the atoms in $z$ direction with linear polarization along $y$ direction.
It is convenient to define positive- and negative-frequency parts of the field.
\begin{equation}
\begin{aligned}
	 E(z)  :=  E_+(z) +  E_-(z) 
\end{aligned}
\end{equation}
where $E_+(z) \simeq i\sqrt{\frac{\hbar\omega_p}{2V\epsilon_0}} \sum_k  a_k e^{i kz}$ and $  E_-(z) =  E_+^\dagger(z)$.
Here ${\omega_p}$ is the central frequency of the quantum dot photon profile (probe).
Since there is no external field, the choice of quantization axis for an atom could be random.
We define the quantization axis also in $y$ direction, same as the photon linear polarization axis. 
Thus the photon will induce $\pi$ transitions of the atoms in this convention.
In the Schr\"odinger picture, the Hamiltonian of all the atoms located at  different position $z$ interacting with the same field can be written as (the definition of the atomic sublevels is presented in Fig. 1 of the main text).
\begin{equation}
\begin{aligned}		
	 H_N=  \sum_{n=1}^N\hbar\left( \omega_{11}   \sigma_{11}^{(n)} + \omega_{22}  \sigma_{22}^{(n)} + \omega_{33}  \sigma_{33}^{(n)} +\omega_{44}  \sigma_{44}^{(n)} \right) - \sum_{n=1}^N  d_y^{(n)} \left( E_+(z_n) +  E_-(z_n)  \right)+ \sum_k\hbar\omega_{k} a_k^{\dagger}  a_k  
\end{aligned}
\end{equation}
where $ E(z_n)$ is the field value at the position of the $n$th atom.

To move everything into the interaction picture, we first consider the photon field part.
Choose $ H_{ph}':= \sum_k\hbar\omega_{k}  a_k^{\dagger}  a_k$.
Field operators in interaction picture is calculated via unitary operator $ U_{ph}(t)=e^{-i  H_{ph}' t/\hbar }$ and can be expressed as $E_+(z, t) = U^\dagger_{ph}(t)  E_+(z)  U_{ph}(t) := {\mathcal{E}}(z,t) e^{-i(\omega_p t - k_p z)},\quad  E_-(z, t) =  E_+^\dagger(z, t)$
where $k_p:=\omega_p/c$ is the central wave vector and a slowly varying envelope is assumed to be
\begin{equation}
\begin{aligned}
	{\mathcal{E}}(z,t) :\simeq i\sqrt{\frac{\hbar\omega_p}{2V\epsilon_0}} \sum_k  a_ke^{-i(\delta\omega_{k}\,t- \delta k\,z)}
\end{aligned}
\end{equation}
where $\delta\omega_{k} := \omega_{k}-\omega_{p}$, and $\delta k := k-k_p$.
For the picture transformation for the $n$th atom we adopt a spatially rotating unitary operator $U_{atom}^{(n)}(z,t)=e^{-\frac{i}{\hbar}H_{atom}'^{(n)}(t-\frac{z_n}{c}) }$, with unperturbed atomic Hamiltonian $H_{atom}'^{(n)}:= \hbar\omega_p\left(  \sigma_{33}^{(n)} + \sigma_{44}^{(n)} \right)$.
Thus the picture transformation operator for the whole system could be defined as
\begin{equation}
\begin{aligned}
	U_N(t) = \prod_{n=1}^N\otimes \left[ \left(\sigma_{11}^{(n)} +  \sigma_{22}^{(n)}\right) + e^{-i(\omega_p t-k_p z_n)} \left(\sigma_{33}^{(n)}  +  \sigma_{44}^{(n)}\right) \right]\otimes e^{-i\sum_k\omega_{k} a_k^{\dagger} a_k t} 	 
\end{aligned}
\end{equation}
The dipole moment operator is   $d^{(n)}_y(t)=\left( d_y^{(31)}\sigma_{31}^{(n)} +d_y^{(41)}\sigma_{41}^{(n)} +d_y^{(32)}\sigma_{32}^{(n)} + d_y^{(42)}\sigma_{42}^{(n)}\right) e^{i(\omega_p t - k_p z_n)}  + h.c.$ in the interaction picture.
Similarly, the Hamiltonian of the whole system is defined by $H_N(t) = U_N^\dagger(t) \left( H_N - \sum_nH_{atom}'^{(n)}-H'_{ph} \right) U_N(t)$, from which one gets
\begin{equation}\label{Eq:HamiltonianI_4Level_generalProbe_NAtom}
\begin{aligned}		
	H_N(t) =&  \sum_{n=1}^N \hbar\left( \omega_{21} \sigma_{22}^{(n)}  - \Delta_{p} \sigma_{33}^{(n)}  +(\omega_{43}-\Delta_p) \sigma_{44}^{(n)}  \right) \\
	 &- \sum_{n=1}^N\left(d_y^{(31)}\sigma_{31}^{(n)}  +d_y^{(41)}\sigma_{41}^{(n)}  +d_y^{(32)}\sigma_{32}^{(n)}  + d_y^{(42)}\sigma_{42}^{(n)} \right)\, \mathcal{E}(z_n,t)+ H.c.
	\end{aligned}
\end{equation}
Here $\Delta_p:=\omega_p-\omega_{31}$ is the detuning between the photon carrier frequency and the $F=1\leftrightarrow F'=1$ transition. $\hat \sigma_{ij}:= \ket{i}\bra{j}$ represent the atomic operators.
We use dipole moment matrix elements \cite{Brink.Satchler:1968} $d_y^{(31)}:=\frac{1}{2\sqrt 3} d_\text{fine} $, $d_y^{(41)}:=\frac{\sqrt{5}}{2\sqrt 3} d_\text{fine} $, $d_y^{(32)}:=\frac{\sqrt{5}}{2\sqrt 3} d_\text{fine} $ and $d_y^{(42)}:=\frac{\sqrt{5}}{2\sqrt 3} d_\text{fine} $, with $d_\text{fine} = 3.588\times 10^{-29}$ C$\cdot$m being the fine transition dipole moment of D1 line. 
The continuous limit of the Hamiltonian is
\begin{equation}\label{Eq:HamiltonianI_4Level_generalProbe_continuous}
\begin{aligned}		
	H(z,t) =&  \hbar\, \omega_{21} \sigma_{22}  - \hbar\, \Delta_{p} \sigma_{33}  + \hbar\, (\omega_{43}-\Delta_p) \sigma_{44}  \\
	& -  \left(d_y^{(31)}\sigma_{31}  +d_y^{(41)}\sigma_{41}  +d_y^{(32)}\sigma_{32} + d_y^{(42)}\sigma_{42} \right)\, \mathcal{E}(z,t) + H.c.
	\end{aligned}
\end{equation}
In the above expression, a position-independent term like $\omega_{31}\sigma_{22}$ means that this term is a common property that any atom in the ensemble has regardless of its position, while a position-dependent term like $\mathcal{E}(z,t)$ gives specific value for an atom at that position.
The many-body Hamiltonian in Eq.\eqref{Eq:HamiltonianI_4Level_generalProbe_NAtom} is thus mapped into its continuous limit in Eq.\eqref{Eq:HamiltonianI_4Level_generalProbe_continuous}.
Similarly, the dipole moment operator 
in continuous limit becomes \\$d_y(z,t)=\left( d_y^{(31)}\sigma_{31} +d_y^{(41)}\sigma_{41} +d_y^{(32)}\sigma_{32} + d_y^{(42)}\sigma_{42}\right) e^{i(\omega_p t - k_p z)}  + H.c.	 $.

The time evolution of atomic ensemble is governed by the Lindblad master equation \cite{Fleischhauer.Imamoglu:2005,Breuer.Petruccione:2007}:
\begin{equation}\label{Eq:master}
    \frac{\partial{ \rho(z,t)}}{\partial t} = -\frac{i}{\hbar}	 [H(z,t), \rho(z,t)] + \sum_{eg}\frac{\Gamma_{eg}}{2} \left(2 \sigma_{ge} \rho(z,t)\, \sigma_{eg} - \sigma_{ee} \rho(z,t) - \rho(z,t)\,  \sigma_{ee} \right)  
\end{equation}
where $e=3,4$ are excited states, $g=1,2$ are ground states.
 We use the atomic polarization decay rates $(\Gamma_{31},\Gamma_{32},\Gamma_{41},\Gamma_{42})=(\frac{1}{6}, \frac{5}{6}, \frac{1}{2},\frac{1}{2})\, \Gamma$, and the spontaneous emission rate $\Gamma = 2\pi\times5.746 \text{ MHz}$.


Meanwhile, the propagation of the single photon field slow envelope is governed by the Maxwell's equation
\begin{equation}
\begin{aligned}
	\left( \nabla^2 -\frac{1}{c^2} \frac{\partial^2}{\partial t^2}\right) E(z,t) &= & \mu_0 \frac{\partial^2 }{\partial t^2}  P_y (z,t)
\end{aligned}
\end{equation}
The left-hand-side (LHS) of the equation can be expressed as $i \frac{2\omega_p}{c^2} \left( \frac{\partial {\mathcal E}(z,t)}{\partial t} +c\frac{\partial {\mathcal E}(z,t)}{\partial z}\right) e^{-i (\omega_p t - k_p z)}  +H.c.$
in the slow envelope approximation.
The right-hand-side (RHS) of the equation can also be simplified noticing that expectation of dipole moment operator is \\$\braket{ P_y(z,t)} :=n\, \text{Tr}[{d_y}(z,t)\,\rho(z,t) ] =n\left(d_y^{(31)}\rho_{13}(z,t)+ d_y^{(32)}\rho_{23}(z,t) + d_y^{(41)}\rho_{14}(z,t)+d_y^{(42)}\rho_{24}(z,t)\right)e^{i(\omega_p t - k_p z)}  + H.c. $.
After comparing the same oscillating terms on both the LHS and RHS of Maxwell's equation, one gets first-order propagation of the single photon field
\begin{equation}\label{Eq:propagation_4Level_continuous}
\begin{aligned}
	\left(\frac{\partial }{\partial t} +c\frac{\partial }{\partial z}\right) {\mathcal E}(z,t)  = i\frac{\omega_p n\, }{2\epsilon_0}\left[d_y^{(13)}\rho_{31}(z,t)+ d_y^{(23)}\rho_{32}(z,t) + d_y^{(14)}\rho_{41}(z,t)+d_y^{(24)}\rho_{42}(z,t)\right] 
\end{aligned}
\end{equation}

We use a Matlab routine to solve the coupled set of Maxwell-Bloch equations (Lindbladian master equation \eqref{Eq:master} and the first order Maxwell's equation \eqref{Eq:propagation_4Level_continuous}), using as input the fitted effective Gaussian distribution in time $\mathcal{E}_\text{eff} (t)$ and obtained all the elements of the atomic density matrix. This provides full knowledge of the dynamics involved in the light-matter interface.

\section{Comparison of Simulation with Experimental Data}
So far, our simulation has only taken into account the frequency spread of the quantum dot photon. However, we have not considered the time uncertainty in the emission of the photons, which is provided by the Wigner-Weisskopf theory \cite{Weisskopf.Wigner.1930} applied to the bi-exciton transition. To characterize the time uncertainty in the emission of the photon, we excite the quantum dot with a pulsed laser at an 80 MHz repetition rate. After cross-polarization filtering and transmission spectrometer, only the bi-exciton photons are collected and time-tagged in SNSPD. Here we emphasize that there is no atomic interaction before the measurement.

\begin{figure}[htb]
\includegraphics[width=.5\textwidth]
{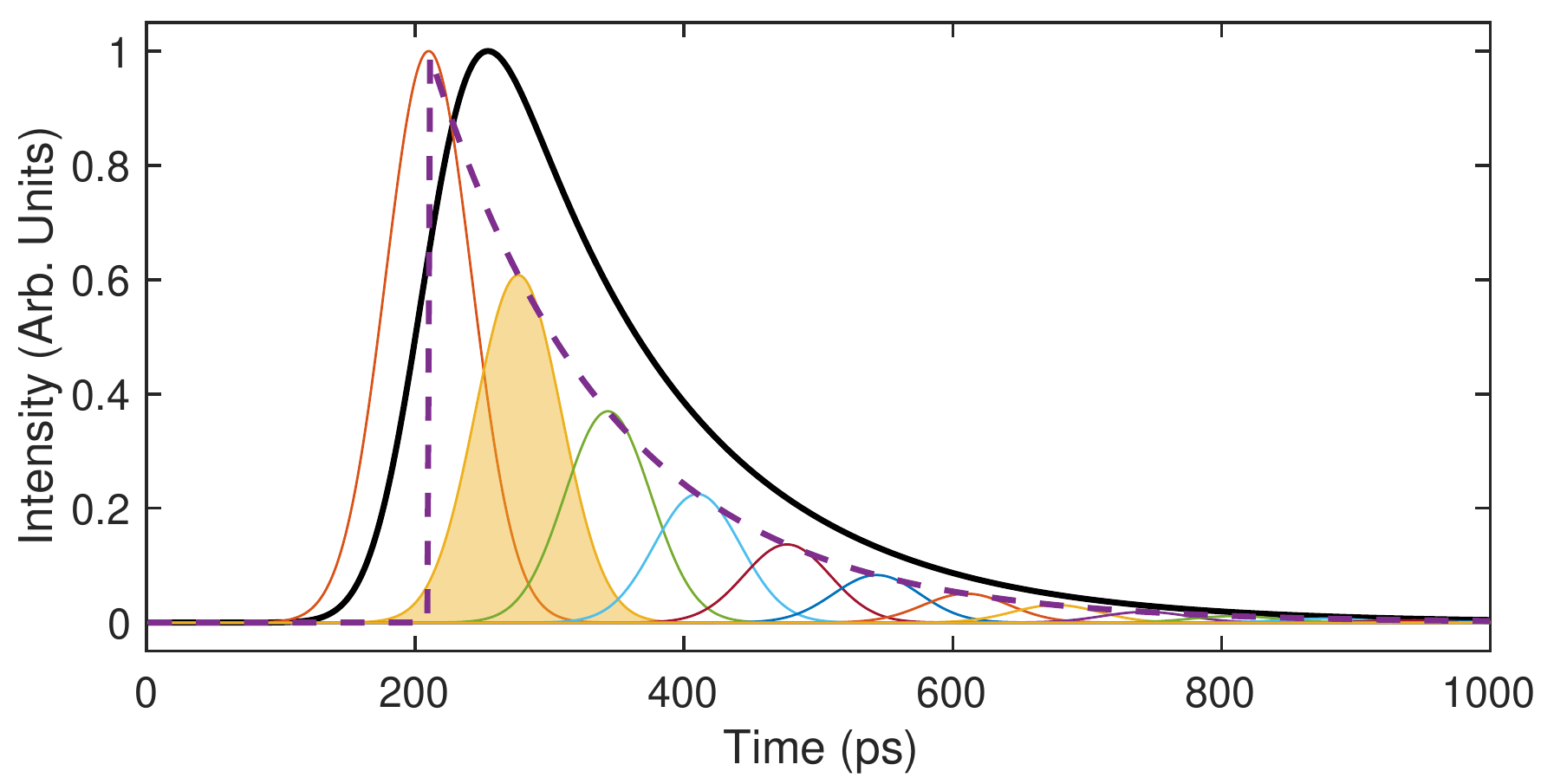}
\caption{
The temporal profile of the quantum dot photon. The black line shows the EMG fitting of the measured photon temporal histograms from an SNSPD measurement. The EMG is the convolution of an exponential decay profile (dashed purple line, with a fitted lifetime of 134 ps) and a Gaussian function (red solid line, FWHM 77.56 ps). An example photon temporal profile applied to the simulation is marked in the yellow area plot.
}
\label{fig:temporal_profile_decomposition}
\end{figure}

Fig. \ref{fig:temporal_profile_decomposition} shows the measured temporal envelop of the quantum dot photons (solid black line). We fit the measured histogram using an exponentially modulated Gaussian function (EMG)
\begin{equation}
    emg(t;t_0, \Delta{t}, \tau) := \frac{1}{2\tau} \exp{\left(-\frac{t- t_0}{\tau} + \frac{\Delta{t}^2}{16\ln(2)\, \tau^2} \right)} ~\text{erfc}\left(-2\sqrt{\ln(2)}\frac{t-t_0}{\Delta{t}} + \frac{\Delta t}{4\sqrt{\ln(2)\tau}}  \right)
\end{equation}
The EMG represents the convolution of the effective Gaussian temporal envelope of the photon together with a time uncertainty that has an exponential decay form, with a fitted lifetime $\tau = 134$ ps.
The EMG-fitted Gaussian temporal profile has a FWHM of $\Delta t=75.35$ ps, which is in good agreement with the 77.56 ps derived from the spectrum discussion in Section \ref{section:envelope}.
The colored Gaussians in Fig. \ref{fig:temporal_profile_decomposition} represent photons arriving at different times weighted by the exponentially decaying probability.

\begin{figure}[htb]
\includegraphics[width = .75 \columnwidth]{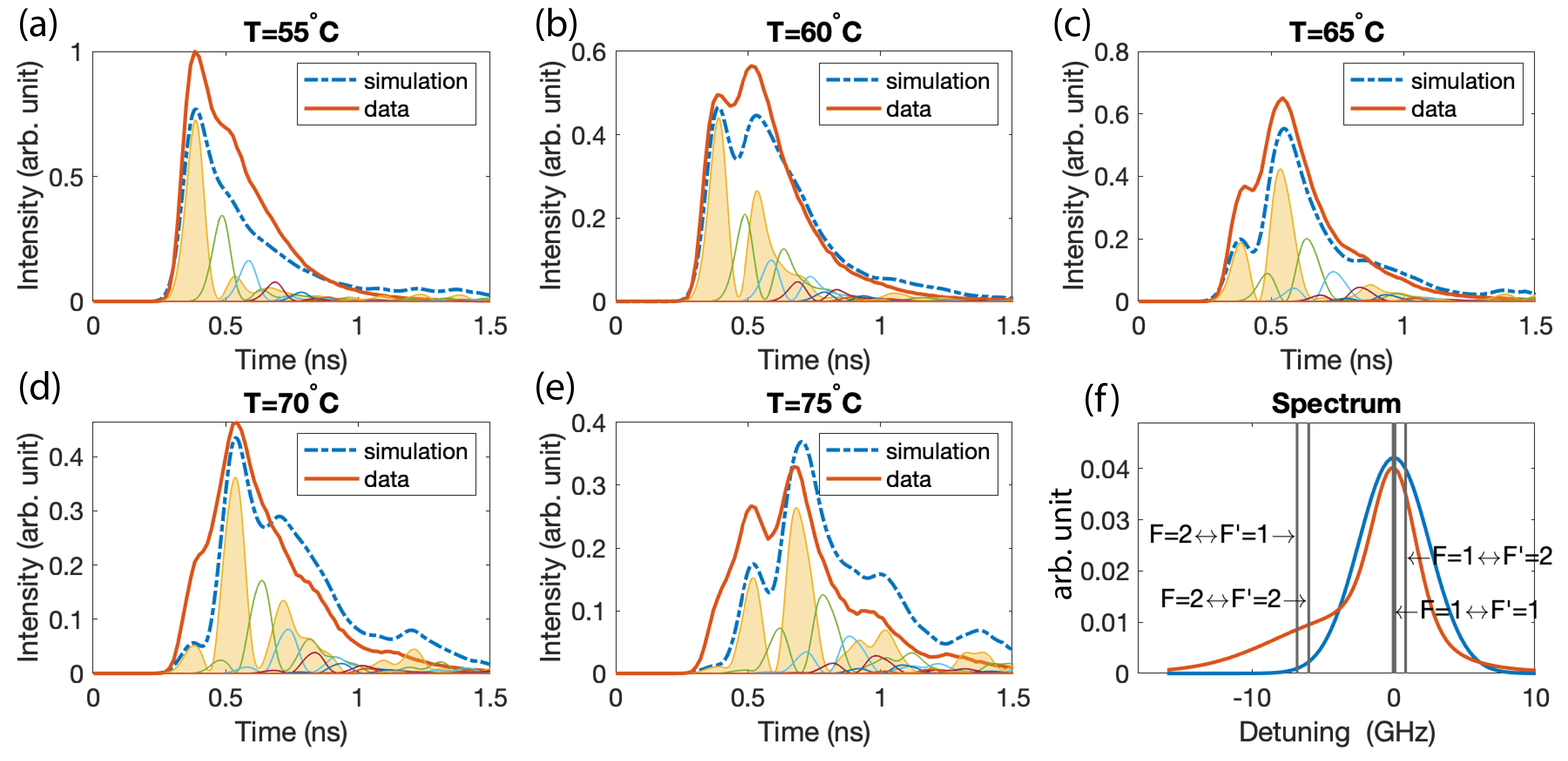}
\caption{(a-e) Benchmarking the convoluted simulations (thick dashed blue lines) against the experimental data (thick red lines) at five temperatures.
The simulated dynamics of the photon envelops reaching the ensemble at different times are represented by the individual thin lines, one of which is highlighted in the yellow area for each temperature.
The individual simulations are weighted by the 134 ps exponential decay function following their arrival times, then integrated to form the convoluted simulation. 
All the simulations are conducted using an identical set of parameters except for ensemble temperatures.
The incident photon field amplitude is the only free global parameter and is optimized for the multi-temperature fit between the simulations and the data.
The raw data are normalized to the $55^{\circ}$C peak.
(f) The red line represents the measured photon spectrum. The effective Gaussian spectrum used in our simulation is shown in the blue line.
The horizontal axis represents the detuning from $^{87} \text{Rb}~D_1( F=1 \leftrightarrow F'=1)$ line.
}
\label{fig:comparison_raw}
\end{figure}

We use these fits to account for the time uncertainty in the measured histograms obtained after transmitting the photons through the atomic ensemble (thick red lines in Fig. \ref{fig:comparison_raw} (a-e)). This is done by convoluting the simulated dynamics (colored thin lines) with the fitted 134 ps exponential decay profile mentioned above. The results of this procedure are shown in the dashed blue lines for different ensemble temperatures. These results match well with the measured histograms, justifying the procedure. Furthermore, we mention that all the parameters remain identical among the simulations (defined as global parameters), except for different ensemble temperatures (defined as local parameters).
The amplitude of the original photon envelope is the only free global parameter, which is varied to obtain a multi-temperature fit that best describes the experimental data globally. 

\begin{figure*}[hbt]
	\includegraphics[width = .85\textwidth]
 {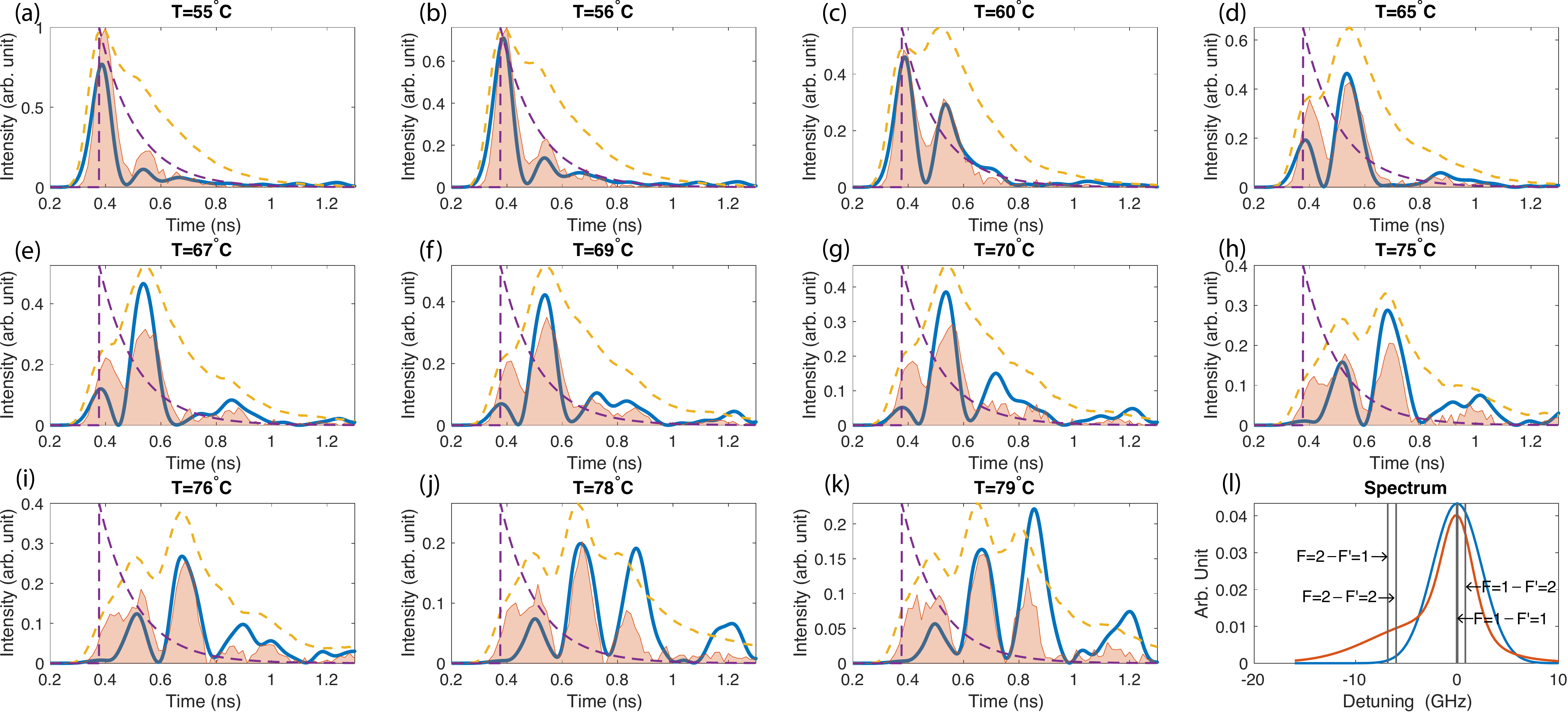}
	\caption{(a-k) Benchmarking the simulations (blue lines) against the deconvoluted experimental data (red areas). 
	The deconvoluted data at each temperature is obtained from the measured histogram (dashed yellow line) deconvoluted with the exponential decay (dashed purple line), and is normalized to the 55$^\circ$C peak.
    The simulations are done with the same set of global parameters for different temperatures. We use an effective Gaussian temporal profile with 77.56 ps FWHM as the input field. The field amplitude is the only free global parameter that is varied to fit all the simulations with the data simultaneously.
	(l) The spectrum of the simulated effective photon is shown in the blue curve and the measured one in the red line.}
	\label{fig:tempseries_11}
\end{figure*}

In Fig. \ref{fig:tempseries_11}, we show the inverse to the procedure described in Fig. \ref{fig:comparison_raw}. We compare the simulated photon dynamics (blue lines) against the deconvoluted experimental histograms (red areas).  These histograms are obtained from the directly measured histograms (dashed yellow lines), deconvoluted with the 134 ps exponential decay profile (dashed purple lines), in order to remove the time uncertainty of the spontaneous emission. Thus they correspond to the photon temporal envelopes reaching the ensemble at a well-defined time, which is the main input to the simulation. This is the procedure use to obtain the figures in the main manuscript. The global agreement between the numerical simulation and the experimental data enables us to make inferences about the various physical phenomena mentioned in the main text. 

\section{Ensemble energy accumulation}
\begin{figure}[htb]
	\includegraphics[width = .7\columnwidth]{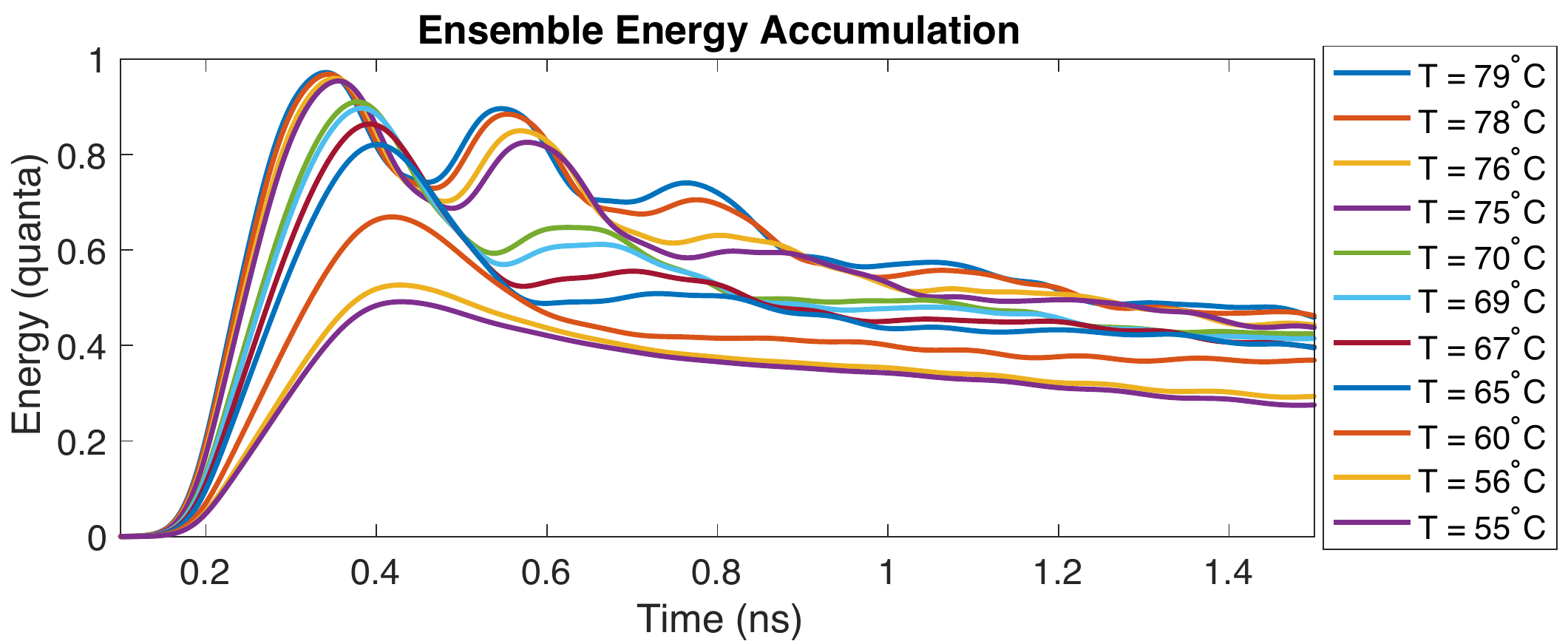} 
	\caption{Calculated ensemble energy accumulation using the simulated atomic density matrix elements during and after the quantum dot photon passes through the ensemble.
    Here we follow the definition of the relative ensemble energy in Eq.\eqref{Eq:ensemble_energy}.
The calculation covers the 11 temperatures for which experiments have been done.
The energy gain is defined in units of $\hbar\omega_p$.
It can be shown that part of the photon energy is retained in the atomic ensemble after it propagates through the cell. 
And this storage time is more than 20 times longer than the effective photon time width.
	}
	\label{fig:ensemble_energy_11}
\end{figure}

Once the atomic density matrix elements are obtained in our simulation, we calculate the additional ensemble energy gained by the atomic ensemble after interacting with the photons. We choose the ground state $\ket{1}$ as a reference and define the total ensemble energy relative to that. The energy stored in the atomic ensemble at a specific time $t$ is then defined as 
\begin{equation}
\begin{aligned}\label{Eq:ensemble_energy}
	\text{Energy}(t) = \sum_{n=1}^N  \hbar\left(  \omega_{21} \rho_{22}^{(n)} (t) + \omega_{31} \rho_{33}^{(n)}(t) + \omega_{41} \rho_{44}^{(n)} (t)\right)
\end{aligned}
\end{equation}
where $N$ is the total number of atoms inside the crossed volume of the photon path with the atomic ensemble, and $\rho_{ii}^{(n)}(t)\, (i=2,3,4)$ are the diagonal density matrix elements of the $nth$ atom at a specific time $t$. For convenience, the energy could also be defined in units of the photon energy (quanta) $\hbar\omega_p$. We use this definition in Fig. 5 of the main text that corresponds to $75^\circ$C, as well as Fig. \ref{fig:ensemble_energy_11} that corresponds to all the eleven temperatures.

\begin{figure}[htb]
 \includegraphics[width = 0.5\columnwidth]{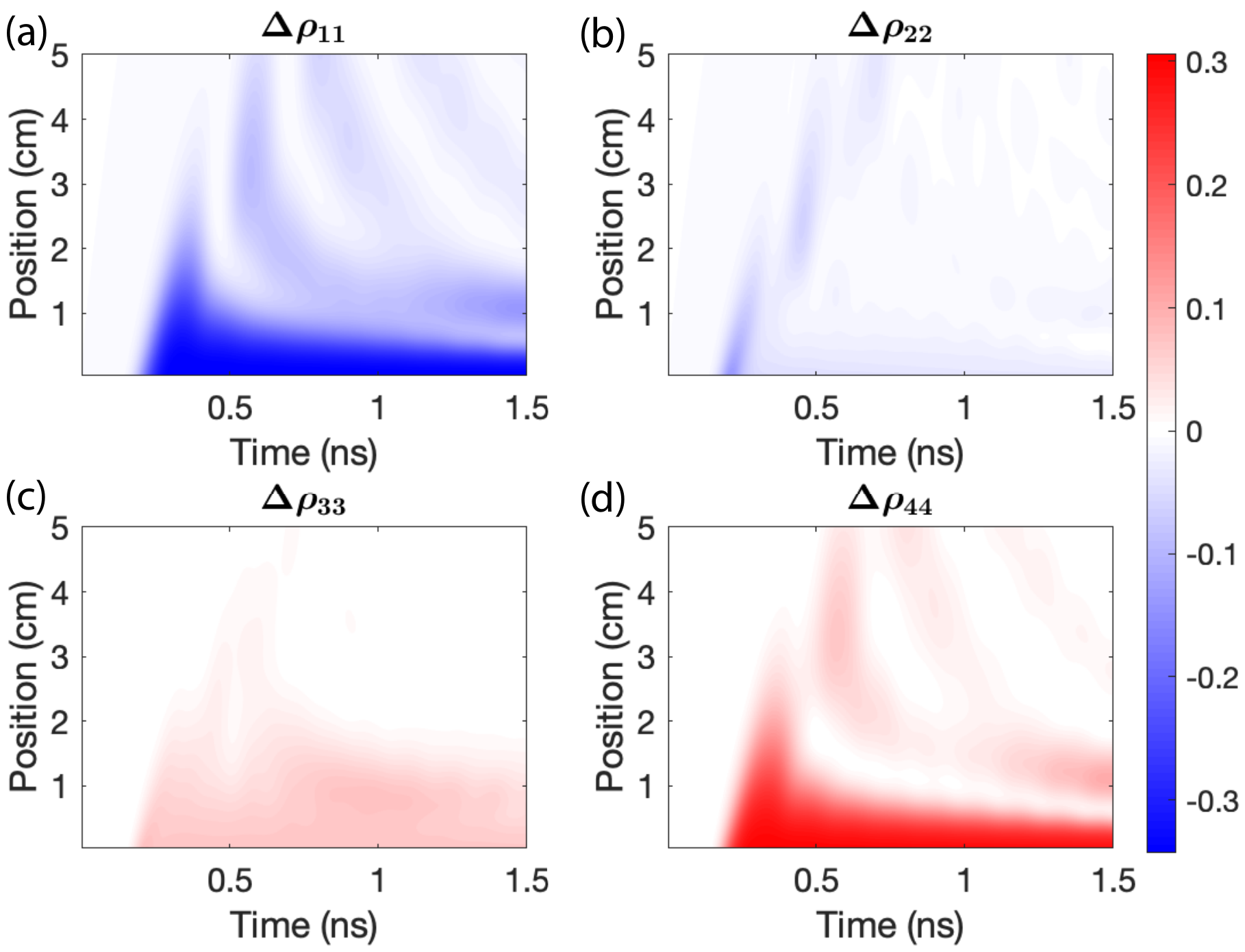}
 \caption{ 
Simulated spatio-temporal evolution of the diagonal density matrix elements that represent population evolution during- and after the interaction with quantum dot photon when the cell temperature is at 75 $^\circ{C}$.
(a-b) Population drop in the ground states $\ket{F=1(2)}$ of $D_1$ transition compared with their thermal equilibrium values before the photon enters. (c-d) Population increase in the excited states $\ket{F'=1(2)}$.
}
	\label{fig:atomic_dynamics_75}
\end{figure}

As the energy accumulation is a function of the atomic populations, we show in Fig. \ref{fig:atomic_dynamics_75} the simulated spatio-temporal evolution of the atomic populations for the four levels as a function of time and propagation depth at 75 $^\circ{C}$.

\bibliography{reference}
\bibliographystyle{apsrev4-1}